\documentclass[12pt,preprint]{aastex}
\newcommand{\mincir}{\raise
-3.truept\hbox{\rlap{\hbox{$\sim$}}\raise4.truept\hbox{$<$}\ }}
\newcommand{\magcir}{\raise
-3.truept\hbox{\rlap{\hbox{$\sim$}}\raise4.truept\hbox{$>$}\ }}
\newcommand{\minmag}{\raise
-3.truept\hbox{\rlap{\hbox{$<$}}\raise5.truept\hbox{$<$}\ }}
\newcommand{\be}{\begin{equation}}
\newcommand{\ee}{\end{equation}}
\newcommand{\ba}{\begin{eqnarray}}
\newcommand{\ea}{\end{eqnarray}}
\newcommand{\brr}{\begin{array}}
\newcommand{\err}{\end{array}}
\newcommand{\bc}{\begin{center}}
\newcommand{\ec}{\end{center}}

\shorttitle{Cluster formation rate in 
models with ``dark energy''}
\shortauthors{Basilakos}

\begin{document}

\title{Cluster formation rate in models with ``dark energy''}

\author{Spyros Basilakos}
\affil{Institute of Astronomy \& Astrophysics, National Observatory of
Athens, I.Metaxa \& B.Pavlou, P.Penteli 152 36, Athens, Greece}

\section*{Abstract}
Based on flat 
Friedmann-Robertson Walker cold dark matter (CDM) 
type models driven by non-relativistic matter and an
exotic fluid (quintessence) with an equation of state: 
$p_{Q}=w\rho_{Q}$ ($-1\le w<0$), we investigate whether or not 
the large scale dynamical effects regarding the cluster formation 
and virialization are related to the cosmic equation of state. 
Using the non-linear spherical collapse 
we find that the cluster formation rate, in quintessence models, 
is intermediate between the open and $\Lambda$CDM respectively.    
For the QCDM case, using the virial theorem and energy conservation and 
assuming a spherical mass overdensity shell, 
we obtain analytically the ratio between the final (virial) and 
the turn-around radius. 
We find that the above ratio is almost independent from the equation
of state.  

{\bf Keywords:} clusters: formation history-
cosmology:theory - large-scale structure of universe 

\section{Introduction}

Recent advances in observational cosmology 
have strongly indicated that we are living in a 
flat, accelerating universe with low matter (baryonic and dark matter) density
(Riess et al. 1998; Perlmutter et al. 1999; de Bernadis et al. 2000;  
Efstathiou et al. 2002; Persival et al. 2002 and references therein). 
The available high quality cosmological data (Type Ia supernovae, 
CMB, etc.) are well fitted by an emerging ``standard model'', which
contains cold dark matter (CDM) to explain clustering and an extra component
with negative pressure (dark energy), usually named ``quintessence'',  
to explain the inflationary flatness prediction ($\Omega_{\rm tot}=1$) as well.

The last few years there have been many theoretical speculations regarding the nature 
of the above exotic dark energy. Most of the 
authors claim that a time varying $\Lambda$-parameter (cf. Ozer \& Taha 1987), 
a scalar field which rolls down the potential $V(\phi)$ (Ratra \& Peebles 1988;
Frieman et al. 1995 
and references therein) or an extra ``matter'' component, which is described by an equation of state
$p_{Q}=w\rho_{Q}$ where $-1\le w <0$, could be 
possible candidates for quintessence. As a particular case, $\Lambda$-models can be 
obtained from quintessence models for $w=-1$.
An excellent paper by Tsagas (2001) gave a 
different perspective to the problem claiming that  
maybe the effects of magnetic fields could 
resemble that of dark energy through the coupling between 
magnetism and curvature of spacetime.
While a variety of observations indicate that $w\le -0.6$ for a flat universe 
(cf. Efstathiou 1999; Dev, Sethi, \& Lohiya 2001; 
Sereno 2002; Ettori, Tozzi \& Rosati 2002 and references therein). 
In this work, we investigate the cluster formation processes 
utilizing the non-linear spherical model for a family of various 
cosmological models 
in order to understand better the theoretical expectations of  	
negative pressure models as well as the variants from the $\Lambda$CDM case.
  
\section{Basic Equations}
For homogeneous and isotropic cosmologies, driven by 
non relativistic matter and an
exotic fluid (quintessence models) with equation of state, $p_{Q}=w\rho_{Q}$
with $-1 \le w <0$, the Einstein field equations can be 
given by:

\be\label{eq:11}
\left( \frac{\dot{\alpha}}{\alpha} \right)^{2}=
\frac{8\pi G}{3}(\rho_{\rm m}+\rho_{Q})-\frac{k}{\alpha^{2}}
\ee
and 
\be
\frac{\ddot{\alpha}}{\alpha}=-4\pi G[(w+\frac{1}{3})\rho_{Q}+\frac{1}{3}\rho_{\rm m}] \;\;,
\ee
where $\alpha(t)$ is the scale factor, $\rho_{\rm m} \propto (1+z)^{3}$ is the matter density and 
$\rho_{Q} \propto (1+z)^{\beta}$ is the dark energy density [$\beta=3(1+w)$]. While $k=-1, 0$ or 1 for
open, flat and closed universe respectively. Thus, the scale factor evolves according to Friedmann equation:
$H^{2}\equiv (\dot{\alpha}/\alpha)^{2}$.
In order to transform the latter equation from time to redshift 
we utilize the following expression:
\be\label{eq:5}
\frac{dt}{dz}=-\frac{1}{H_{\circ}E(z)(1+z)} \;\;,
\ee
with
\be\label{eq:4}
E(z)=\left[ \Omega_{\rm m}(1+z)^{3}+\Omega_{k}(1+z)^{2}+
\Omega_{Qo}(1+z)^{\beta}\right]^{1/2}
\ee
where the Hubble parameter is given by: $H(z)= H_{\circ} E(z)$,
while $\Omega_{\rm m}= 8\pi G \rho_{o}/3H_{o}^{2}$ 
(density parameter), $\Omega_{k}=-k/H_{\circ}\alpha_{o}$ (curvature parameter),
$\Omega_{Qo}= 8\pi G \rho_{Qo}/3H_{o}^{2}$ 
(dark energy parameter) at the present time, which satisfy
$\Omega_{\rm m}+\Omega_{k}+\Omega_{Qo}=1$
and $H_{\circ}$ is the Hubble constant. 

In addition, to $\Omega_{m}(z)$ also $\Omega_{Q}(z)$ could evolve 
with redshift as
\be
\Omega_{\rm m}(z)=\frac{\Omega_{\rm m} (1+z)^{3}}{E^{2}(z)} \;\;\;{\rm and} \;\;\; 
\Omega_{Q}(z)=\frac{\Omega_{Qo} (1+z)^{\beta}}{E^{2}(z)} \;\; .
\ee
Note, that $\Lambda$-models can be described by quintessence models with $w$ strictly equal to -1
while if $w=0$ the equation of state behaves like that of pressureless matter.
It is interesting to mention that different values of 
$w$ could yield flat ($\Omega_{k}=0$) cosmological models 
for which there is not a one-to-one correspondence between the global geometry and
the expansion of the universe. Indeed, in flat low-$\Omega_{\rm m}$ with $w=-1/3$ model, 
the functional form of the dark energy density can be given by 
$\rho_{Q} \propto (1+z)^{2}$.
Therefore, the equation of state $p_{Q}=w\rho_{Q}$ 
leads to the same expansion as in an open universe.
In other words, $p_{Q}$ plays a similar role to the curvature, despite 
the fact that this quintessence model has a spatially flat geometry!
Also in the case of a flat low-$\Omega_{\rm m}$ with $-1<w<-1/3$ or $0< \beta <2$ model, 
we have $\rho_{Q} \propto \alpha^{-\beta} 
< \alpha^{-3} \propto \rho_{\rm m}$, which means that 
the dark energy density falls off at a slower rate than cold dark matter.
This is very important because the dark energy component starts to 
dominate the mass density
in the universe, especially at the late times, thus creating an 
accelerating expansion.

As for the power spectrum of our CDM models, 
we consider $P(k) \approx k^{n}T^{2}(k)$ with
scale-invariant ($n=1$) primeval inflationary fluctuations. 
In particular, we consider three different spatially flat low-$\Omega_{\rm m}$ 
cosmological models with negative pressure, 
where: 
(1) an $\Omega_{\rm m}=1-\Omega_{\Lambda}=0.3$ with $w=-1$ ($\Lambda$CDM), 
(2) a model with $\Omega_{\rm m}=1-\Omega_{Qo}=0.3$ with $w=-2/3$ (QCDM1) and to this end 
(3) an $\Omega_{\rm m}=1-\Omega_{Qo}=0.3$ with $w=-1/3$ (QCDM2). Our models 
have $\Gamma \sim 0.2$, in approximate agreement with the shape parameter estimated from galaxy 
surveys (cf. Maddox et al. 1990) and Hubble constant $H_{\circ}=100h$ km s$^{-1} $Mpc$^{-1}$ 
(where $h=0.65$). Finally, the latter cosmological models 
are normalized to have fluctuation amplitude in 8 $h^{-1}$Mpc scale of
$\sigma_{8}=0.50 \pm0.1 \Omega_{\rm m}^{-\gamma}$ (Wang \& Steinhardt 1998) with 
$\gamma=0.21-0.22w+0.33\Omega_{\rm m}$.

\section{Cluster formation in QCDM Cosmologies}
The growth factor as a function of redshift \footnote{$D(z)=(1+z)^{-1}$ 
for an Einstein-de Sitter universe.} 
for the mass density contrast, $\delta=(\delta\rho_{\rm m}/\rho_{\rm m})$, modeled as a 
pressureless fluid (cf. Peebles 1993; Lokas 2001 and references therein) is:   
\be\label{eq:24}
D(z)=\frac{5\Omega_{\rm m} E(z)}{2}\int^{\infty}_{z} \frac{(1+y)}{E^{3}(y)} \;\;.
{\rm d}y 
\ee
In case of $w=-2/3$, $D(z)$ can be obtained by Wang \& Steinhardt (1998; their
equation 14).
Below we attempt to investigate how the growing mode of perturbations 
is affected by the time evolution of the dark energy
contribution to $\Omega_{\rm tot}$. To do so 
we follow the ideas of Lightman \& Schechter 1990, 
Lahav et al. (1991), and Carrol, 
Press \& Turner (1992), who use a convenient 
approximation for $D(z)$: 
\be\label{eq:25}
D(z)=\frac{5\Omega_{\rm m}}{2(1+z)} 
\left[ \Omega_{\rm m}^{\cal \alpha}-\Omega_{Q}+ 
(1+\frac{1}{2} \Omega_{\rm m}) (1+{\cal A} \Omega_{Q}) \right]^{-1} \;\;,
\ee
where we leave unconstraint the values of $\alpha$ and ${\cal A}$ and
we use the generic expression for ${\cal \alpha}$, 
defined by the Wang \& Steinhardt (1998):
\be\label{eq:26}
{\cal \alpha} \simeq \frac{3}{5-w/(1-w)}+
\frac{3}{125}\frac{(1-w)(1-3w/2)}{(1-6w/5)^{3}}(1-\Omega_{\rm m}) \;\;.  
\ee 
Utilizing a $\chi^{2}$ minimization procedure 
between the complete $D(z)$ solution and equations 
(\ref{eq:25}), (\ref{eq:26}), considering 
different values for ${\cal A}$ and $w$ for $\Omega_{\rm m}=0.3$,
we obtain a roughly quadratic relation for ${\cal A}$, fitted by:
${\cal A}\simeq 1.742+3.343w+1.615w^{2}\;\;.$ 
Note that for the $\Lambda$CDM case ($w=-1$), the result 
${\cal A} \simeq 0.014$ with ${\cal \alpha}\simeq 6/11$ 
agrees with Lahav et al. (1991) who found
${\cal A} \simeq 1/70$ and ${\cal \alpha}\simeq 0.6$.
Finally, for the quintessence models
QCDM1 ($w=-2/3$) and QCDM2 ($w=-1/3$) we have 
$({\cal A},{\cal \alpha}) \simeq (0.232,0.565)$ and    
$({\cal A},{\cal \alpha}) \simeq (0.810,0.584)$ respectively.
In Figure 1, lines represent 
the complete growing mode solution in contrast to 
the $D(z)$ approximation formula, which is represented by the points. 
Thus, it becomes evident that the growing mode 
approximation works extremely well. 

The concept of estimating the fractional rate of cluster formation, 
considering the non-linear spherical collapse model, has 
been brought up by different 
authors (cf. Peebles 1984; Weinberg 1987; 
Martel \& Wasserman 1990; Lahav et al. 1991; 
Richstone, Loeb \& Turner 1992). Here, we present the 
basic steps of the cluster formation processes
that we will use, following the above ideas. Basically, the 
above authors introduced a procedure which computes the 
rate at which mass join virialized structures, which grow from small 
initial perturbations in the universe.
It is well known that the basic cosmological 
equations, mentioned before, are correct either for 
the entire universe or for homogeneous spherical perturbations 
[by replacing the scale factor with radius $R(t)$]. Therefore, having 
bound perturbations 
which do not expand forever the time they need 
to recollapse (at some redshift $z=z_{\rm f}$) 
is twice the turn-around time $t_{\rm f}=2t_{\rm ta}$. 

Furthermore, making the assumption that when the matter 
epoch just begins, the universe 
is described by (i) an unperturbed Hubble flow, 
(ii) a matter fluctuation field 
which has a Gaussian distribution: 
\be\label{eq:88}
dF(\Delta)=\frac{1}{\sqrt{2\pi}\sigma} {\rm exp} \left( -\frac{\Delta^{2}}{2\sigma^{2}} \right) d \Delta \;\;, 
\ee
where $\Delta$ is the mass density contrast
and it contains a dark energy component. 
The rms mass fluctuation amplitude at 
8 $h^{-1}$Mpc can be expressed as a function of redshift as 
$\sigma(z)=D(z)\sigma_{8}$.
Then, for the bound perturbations that we 
consider, following the standard
cluster formation pattern (cf. Richstone et al. 1992), we can 
obtain the ratio of the collapse time to the 
current age of the (unperturbed) universe: 

\be
\frac{t}{T_{\circ}}=2 \int_{z}^{\infty} 
\frac{{\rm d}x}{(1+x) M(\Delta,x)}
\left[\int_{0}^{\infty} \frac{{\rm d}x}{(1+x)E(x)} \right]^{-1}
\ee 
where
$$M(\Delta,z)=[ \Omega_{\rm m}(1+\Delta)(1+z)^{3}+
(1-\Omega_{\rm m}(1+\Delta)-$$
$$\Omega_{Qo})(1+z)^{2}+\Omega_{Qo}(1+z)^{\beta}]^{1/2} \;\;.$$

The solution of the integral of eq.(\ref{eq:88}), which is Gaussian, 
describes the fraction of the universe (characterized by 
$\Omega_{\rm m}$, $\Omega_{Qo}$ and $\sigma_{8}$) on some specific mass scale that has already 
collapsed at time $t$ and is given by (see also Richstone et al. 1992):
\be
F \left( \frac{t}{T_{\circ}} \right)=\frac{1}{2} \left[1-{\rm erf} 
\left( \frac{\Delta(t/T_{\circ})}{\sqrt{2} \sigma} \right) \right] \;\;,
\ee
The next step is to normalize the probability to give the number of clusters which
have already collapsed by the epoch $t$ (cumulative distribution), divided 
by the 
number of clusters which have collapsed at the 
present epoch: 
\be\label{eq:89}
{\cal F}=\frac{F(t/T_{\circ})}{f} \;\;\;{\rm with}\;\;\; 
f=\frac{\langle n \rangle M}{\rho_{\rm c} \Omega_{\rm m}} \;\;  .
\ee
Where $\rho_{\rm c} \simeq 2.78 \times 10^{11} h^{2} M_{\odot}$ Mpc$^{-3}$ is the critical density, $\langle n \rangle$ 
is the number density of clusters which have collapsed prior to the present epoch. The parameter
$\langle n \rangle$ can be defined utilizing the Abell/ACO cluster catalog, 
which is a volume-limited sample within 
$\sim 180-200 h^{-1}$ Mpc and which a general agreed value is: 
$\langle n \rangle \simeq 1.8 \times 10^{-5} h^{3}$ Mpc$^{-3}$. Therefore, 
considering 
virialized clusters
of the mass scale of 
rich Abell clusters, $M \simeq 10^{15} h^{-1} M_{\odot}$, it is a 
routine to 
obtain the ratio of collapsed matter at the present time 
$f(10^{15} M_{\odot})=0.065 \Omega_{\rm m}^{-1}$.   

It is obvious that the above generic of form eq.(\ref{eq:89})
depends on the choice of the background cosmology.
Indeed the relationship between $\Delta$, $z$, 
$t/T_{\circ}$ and $\sigma_{8}$ is different in
different cosmologies (cf. Mo \& White 1996; Magliocchetti et al. 2001), where: 
\be
\frac{\Delta(z)}{\sigma_{8}}=\frac{\delta_{\rm c}}{D(z)\sigma_{8}}=
\frac{\delta_{\rm c}}{\sigma(z)} \;\;.
\ee
The value $\delta_{\rm c}=1.686$ corresponds to the spherical top-hat model
in $\Omega_{\rm m}=1$, but it has been shown that $\delta_{\rm c}$ depends 
only weakly on the cosmology (Eke, Cole \& Frenk 1996).
Considering the three low-$\Omega_{\rm m}$ 
spatially flat models, described in section 2 we can obtain 
complementary predictions also for the 
old ``standard'' CDM model with $\Omega_{\rm m}=1$, $h=0.5$ (SCDM) and the 
open model with $\Omega_{\rm m}=0.3$ and $h=0.65$ (OCDM). The latter two 
cosmological models are normalized by the observed cluster abundance at
zero redshift; $\sigma_{8}=0.55\Omega_{\rm m}^{-0.6}$ (Eke et al. 1996).

In figure 2 we present the behavior of eq.(\ref{eq:89}) as a 
function of redshift and $t/T_{\circ}$.   
The behavior of the cluster formation rate, ${\cal F}$, 
has the expected form, ie. it is a decreasing function of 
redshift. For the high density universe $\Omega_{\rm m}=1$
we found the known behavior (Richstone et al. 1992) in 
which galaxy clusters started to form only very recently 
while in an open or a flat low-density    
universe, clusters should appear to be
formed at $z\simeq 2$ due to the fact that clustering effectively freezes 
at high redshifts. In this framework, we have found that 
the cluster formation rate in quintessence models 
has an intermediate growth between that of an   
open and $\Lambda$CDM models respectively.    
Finally, it is obvious that 
for $w \longrightarrow -1$ the quintessence cluster
formation pattern tends to that of $\Lambda$CDM case, as it should.

\subsection{Cluster virialization with dark energy}
In this section, based on the notations of Lahav et al. (1991),
Wand \& Steinhardt (1998) and Lokas \& Hoffman (2003)
we study the cluster virialization in QCDM models. Here we review only some basic 
concepts of the problem. Assuming a spherical mass overdensity shell, 
utilizing both the
virial theorem $T=-\frac{1}{2}U_{G}+U_{Q}$ and the energy 
conservation $T_{f}+U_{G,f}+U_{Q,f}=U_{G,ta}+U_{Q,ta}$ where,
$T$ is the kinetic energy, $U_{G}$ is the potential energy and $U_{Q}$
is the potential energy associated with quintessence. Therefore,  
we can obtain a cubic equation which relates the ratio between the final (virial) $R_{\rm f}$ and 
the turn-around outer radius $R_{\rm ta}$:
\be\label{eq:81}
2n_{u} \left(\frac{R_{\rm f}}{R_{\rm ta}}\right)^{3}
-(2+n_{t})\left(\frac{R_{\rm f}}{R_{\rm ta}}\right)+1=0 \;\;,
\ee
with an approximate solution (see Wand \& Steinhardt 1998):
$$\frac{R_{\rm f}}{R_{\rm ta}}=\frac{1-n_{u}/2}{2+n_{t}-3n_{u}/2}$$
where $ n_{u}=2 \Omega_{Qo} 
(1+z_{\rm f})^{3(1+w)}/\zeta \Omega_{\rm m} (1+z_{\rm ta})^{3}$, 
$ n_{t}=2 \Omega_{Qo} (1+z_{\rm ta})^{3w}/\zeta \Omega_{\rm m}$
and 
$\zeta\equiv \rho_{\rm cl}(z_{\rm ta})/\rho_{\rm m}(z_{\rm ta})$ 
and for $-1\le w \le 0$ is a weakly $\Omega_{\rm m}$-dependent 
(see Wand \& Steinhardt 1998). 
In this study, we derive analytically the 
exact solution of the above cubic equation, having polynomial 
parameters: $a_{1}=0$, $a_{2}=-(2+n_{t})/2n_{u}$ and $a_{3}=1/2n_{u}$. Indeed, using some 
basic elements from the Algebra (see the Appendix) 
the discriminant of eq.(\ref{eq:81}) is:
$${\cal D}(n_{u},n_{t})=\frac{2(2+n_{t})^{3}-27n_{u}}{4n_{u}^{3}} \;\;.$$
Owing to the fact that we are living in an accelerating universe 
$\Omega_{Qo}>0$ with low matter (baryonic and dark matter) density,
the condition for an overdensity shell to 
turn around is $0<n_{t} \le n_{u} <1$ 
which gives ${\cal D}(n_{u},n_{t})>0$ and 
therefore all roots of the cubic equation are real 
(irreducible case) but one of them 
\be
r_{2}=\frac{R_{\rm f}}{R_{\rm ta}}=-\frac{1}{3}[R+\sqrt{3}M]
\ee
corresponds to expanding shells (for $R$, $M$ and the other parameters 
see the Appendix). Thus, according to the solution described above, figure 3 shows the surface behavior 
of the exact virial solution 
in parametric form with: 
$$r^{1/3}=\left(\frac{27^{2}}{16n_{u}^{2}}+\frac{27{\cal D}}{4}\right)^{1/6}\;\;\;
{\rm and}\;\;\;
\frac{\theta}{3}=\frac{1}{3}{\rm arctan}\left(\frac{2n_{u}\sqrt{3D}}{9}\right)$$
It is interesting to say that the minimal possible 
ratio is $R_{\rm f}/R_{\rm ta} \simeq 0.35$ which is similar
to the $\Lambda$CDM solution derived by Lahav et al. (1991)
for which case we have
$n_{u}=n_{t}=n=\Omega_{\Lambda}/\zeta \Omega_{\rm m}(1+z_{\rm ta})^{3}$, where $\Omega_{\Lambda}$ is the
cosmological constant parameter at the present time. The latter minimal ratio implies that the 
final stage of the cluster virialization is almost independent 
from the equation of state. 

\section{Conclusions}
Using the non-linear spherical collapse model, 
we derive the cluster formation rate as a function of redshift
in CDM models with negative pressure. We verify that the 
formation rate in quintessence models 
($-1<w<0$) is an intermediate case 
between the open and $\Lambda$CDM models 
respectively. Furthermore, we study the cluster virialization in QCDM models and we 
derive the exact solution. The minimal possible 
ratio is $R_{\rm f}/R_{\rm ta} \simeq 0.35$ which is consistent
with the $\Lambda$CDM solution derived by Lahav et al. (1991) which means that the above 
ratio is almost independent from the equation of state.  

\section* {Appendix}
Without wanting to appear too pedagogical, we remind the reader of some 
basic elements of Algebra. Given a cubic equtation: $x^{3}+a_{1} x^{2}+a_{2} x+a_{3}=0$.
Let ${\cal } D$ be the discriminant:
\be
{\cal D}=a_{1}^{2} a_{2}^{2}-4a_{2}^{3}-4a_{1}^{3}a_{3}-27a_{3}^{2}+18a_{1}a_{2}a_{3}
\ee  
and 
$$x_{1}=-a_{1}^{3}+\frac{9}{2}a_{1}a_{2}-\frac{27}{2}a_{3} \;,\;\;\;\;\;\;
x_{2}=-\frac{3\sqrt{3 \cal{D}}}{2}$$
then, the roots of the above equation are:
\be
r_{1}=-\frac{a_{1}}{3}+\frac{1}{3}[q_{1}+q_{2}]
\ee  

\be
r_{2}=-\frac{a_{1}}{3}+\frac{1}{3}[\epsilon^{2} q_{1}+\epsilon q_{2}]
\ee  

\be
r_{3}=-\frac{a_{1}}{3}+\frac{1}{3}[\epsilon q_{1}+\epsilon^{2} q_{2}]
\ee  
where $q_{n}=(x_{1}\pm i x_{2})^{1/3}$ and $\epsilon=\frac{-1+\sqrt{-3}}{2}$.
If ${\cal D}<0$, we have one real root ($r_{1}$) and a pair of complex conjugate roots.
If ${\cal D}=0$, all roots are real and at least two of them are equal.
If ${\cal D}>0$, all roots are real (irreducible case). In that case $r_{1}$, $r_{2}$ and $r_{3}$
can be written:
 \be
r_{1}=-\frac{a_{1}}{3}+\frac{2}{3}R
\ee  

\be
r_{2}=-\frac{a_{1}}{3}-\frac{1}{3}[R+\sqrt{3}M]
\ee  

\be
r_{3}=-\frac{a_{1}}{3}-\frac{1}{3}[R-\sqrt{3}M]
\ee  
where $R$ and $M$ are given by
\be
R=r^{1/3} {\rm cos} (\frac{\theta}{3})\;\;\; {\rm and}\;\;\;
M=r^{1/3} {\rm sin} (\frac{\theta}{3})
\ee
with $r=\sqrt{x_{1}^{2}+x_{2}^{2}}$ and $\theta={\rm Arctan} (x_{2}/x_{1})$.

\section* {Acknowledgements}
I thank Manolis Plionis for helpful comments and suggestions.
Finally, I would like to thank the anonymous referee, for his/her 
possitive comments and useful sugestions.

\begin{figure}
\plotone{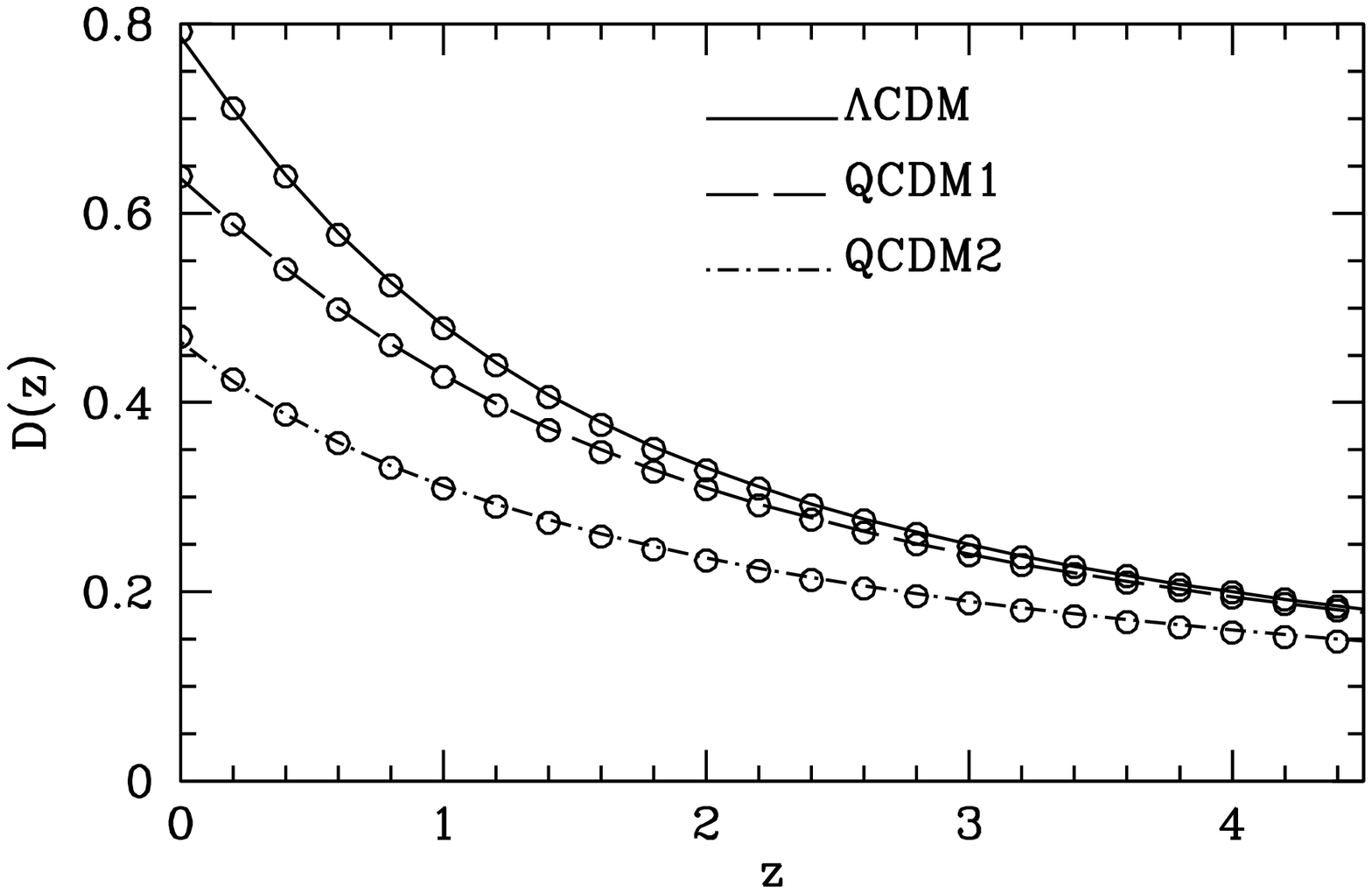}
\caption{Growth factor for matter density fluctuations 
as a function of redshift
in linear theory using cosmological 
models ($\Omega_{\rm m}=0.3$) with negative 
pressure. Note that lines represent 
the complete $D(z)$ solution while points 
represent the $D(z)$ approximation formula.} 
\end{figure}

\begin{figure}
\plotone{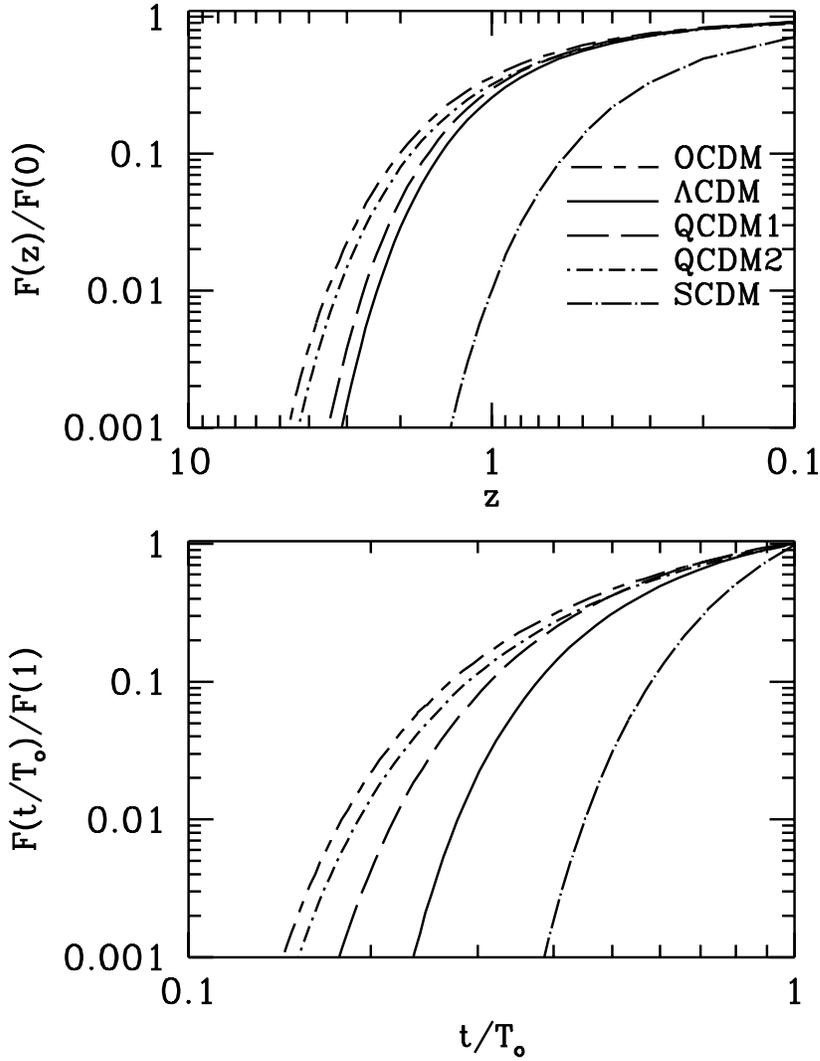}
\caption{Theoretical predictions of the fractional rate of cluster formation
as a function of redshift and fractional time $t/T_{\circ}$, for different cosmological models.}  
\end{figure}

\begin{figure}
\plotone{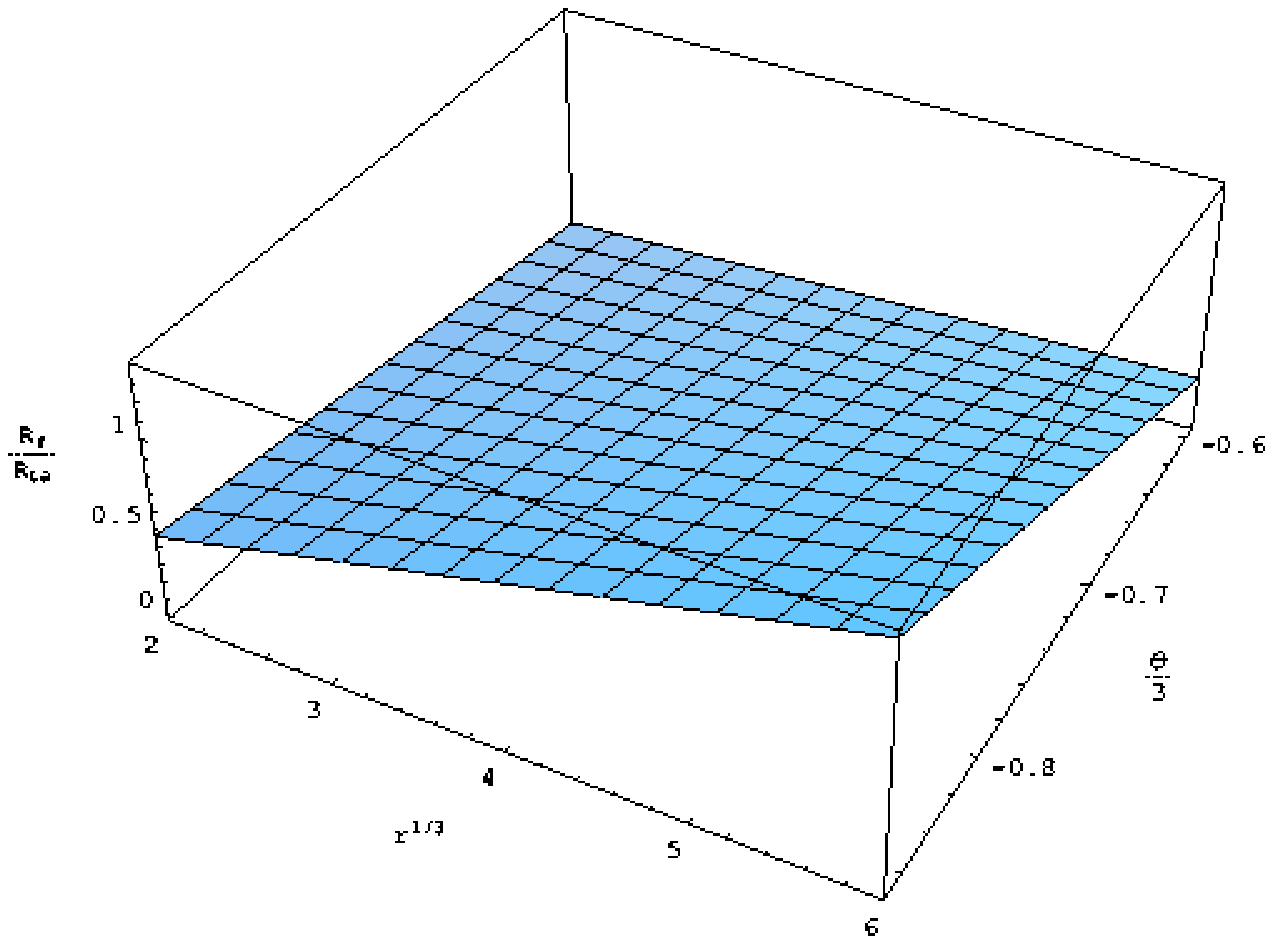}
\caption{The ratio of the final (virial) to turn-around radius of a virialized cluster 
utilizing a QCDM cosmology.We give the surface solution in parametric form ($r^{1/3}$, $\theta/3$
see the Appendix).}  
\end{figure}


{\small  
\begin{thebibliography}{}
\bibitem[]{}Carrol, S. M., Press, W. H., Turner, E. L., Annu. Rev. Stron. Astrophys., 1992, 30, 499
\bibitem[]{}de Bernadis P. et al., 2000, Nature, 404, 955
\bibitem[]{}Dev, A., Sethi, M., Lohiya, D., 2001, PhL, 504, 207
\bibitem[]{}Efstathiou, G., 1999, \mnras, 310, 842
\bibitem[]{}Efstathiou, G., 2002, \mnras, 330, L29
\bibitem[]{}Eke, V., Cole, S., Frenk, C. S., 1996, \mnras, 282, 263
\bibitem[]{}Ettori, S., Tozzi, P, Rosati, P., 2003, A\&A, 398, 879
\bibitem[]{}Frieman, J. A., Hill, C. T., Stebbins A., Waga I., 1995, Phys. Rev. Lett., 75, 2077
\bibitem[]{}Lahav, O., Lilje, P. B., Primack, J. R., Rees, M. J., 1991, \mnras, 251, 128 
\bibitem[]{}Lightman, A. P.\& Schechter, P. L., 1990, \apjs, 74, 831
\bibitem[]{}Lokas, L., E., 2001, Acta Phys.Polon., B32, 3643-3654
\bibitem[]{}Lokas, L., E., \&, Hoffman, Y., 2003, \mnras, {\rm in press}, 
astro-ph/018283
\bibitem[]{}Maddox, S., Efstathiou, G., Sutherland, W. J., Loveday, J., 1990, \mnras, 242, 457
\bibitem[]{}Martel, H., \&, Wassermsn, I., 1990, ApJ, 348, 1
\bibitem[]{}Magliocchetti, M., Moscardini, L., Panuzzo, P., Granato, L. G., De Zotti, G., 2001, 
\mnras, 325, 1553
\bibitem[]{}Mo, H.J, \& White, S.D.M  1996, \mnras, 282, 347
\bibitem[]{}Ozer M., \&, Taha, O., 1987, Nucl. Phys., B287, 776
\bibitem[]{}Peebles P.J.E., 1984, \apj, 284, 439
\bibitem[]{}Peebles P.J.E., 1993. Principles of Physical Cosmology, 
Princeton University Press, Princeton New Jersey
\bibitem[]{}Perlmutter, S., et al., 1999, \apj, 517, 565
\bibitem[]{}Percival, J., W., et al., 2002, \mnras, 337, 1068
\bibitem[]{}Ratra, B., \&, Peebles P. J. E., 1988, Phys. Rev. D, 37, 3406
\bibitem[]{}Richstone, D., Loeb, A., Turner, E. L., 1992, \apj, 393, 477
\bibitem[]{}Riess, A. G., et al., 1998, AJ, 116, 1009
\bibitem[]{}Sereno, M., 2002, A\&A, 393, 757
\bibitem[]{}Tsagas, C., 2001, Phys. Rev. Lett., 86, 5421
\bibitem[]{}Wang, L. \& Steinhardt, P.J., 1998, ApJ, 508, 483
\bibitem[]{}Weinberg, S., 1987, Phys. Rev. Lett., 59, 2607

\end{thebibliography}
}
\end{document}